\documentclass[prl,twocolumn, final,superscriptaddress]{revtex4}
\usepackage{amsmath}
\usepackage{epsfig}
\usepackage{rotating}

\begin{document}
\title{Entanglement and the nonlinear elastic behavior of forests of coiled 
carbon nanotubes}

\author{V. R. Coluci}\email[\footnotesize{Author to whom correspondence 
should be addressed. FAX:+55-19-35215376. Electronic address: }]
{coluci@ifi.unicamp.br}
\affiliation{Instituto de F\'{\i}sica ``Gleb Wataghin",
             Universidade Estadual de Campinas, C.P. 6165, 13083-970
             Campinas SP, Brazil}

\author{A. F. Fonseca}
\affiliation{The MacDiarmid NanoTech Institute, University of Texas, Richardson, 
TX 75083-0688, USA}

\author{D. S. Galv\~ao}
\affiliation{Instituto de F\'{\i}sica ``Gleb Wataghin",
             Universidade Estadual de Campinas, C.P. 6165, 13083-970 Campinas
             SP, Brazil}
 
\author{C. Daraio}
\affiliation{Aeronautics and Applied Physics, California Institute of 
Technology, Pasadena CA, 91125}

\date{\today}

\begin{abstract}

Helical or coiled nanostructures have been object of intense experimental and theoretical studies due to their special electronic and mechanical properties. Recently, it was experimentally reported that the dynamical response of foamlike forest of coiled carbon nanotubes under mechanical impact exhibits a nonlinear, non-Hertzian behavior, with no trace of plastic deformation. The physical origin of this unusual behavior is not yet fully understood. In this work, based on analytical models, we show that the entanglement among neighboring coils in the superior part of the forest surface must be taken into account for a full description of the strongly nonlinear behavior of the impact response of a drop-ball onto a forest of coiled carbon nanotubes.

\end{abstract}


\maketitle

The study of nanostructures, in special carbon nanotubes (CNTs) and nanowires have been object of intense experimental and theoretical investigations due to their large range of possible applications and new physical phenomena \cite{iijima,ray}. Among these nanostructures helical and coiled structures have a special place due to their differentiated mechanical behavior \cite{alexandre}. Coiled carbon nanotubes (CCNTs) were first predicted to exist in the early 1990s by Dunlap \cite{dunlap} and Ihara \textit{et al.} \cite{ihara1,ihara2,ihara3} and experimentally observed in 1994 by Zhang \textit{et al.} \cite{zhang}. CCNTs have been receiving increasing interest because of their additional capability to serve as nanoscale mechanical springs~\cite{chen}, electrical inductors~\cite{motojima}, and for their potential applications in composites~\cite{lau}. 

Recently, the dynamical response of a foamlike forest of CCNTs (Fig. 1(a)) under impact of a drop-ball has been reported~\cite{4}. The experiment consisted of producing arrays of bundles of CCNTs~\cite{wang,bandaru}, let a stainless steel bead falls down on the forest of CCNTs, and measure the dynamic force at the wall below the forest during the stages of penetration and restitution. The analysis of the forest's morphology after impact has shown no trace of plastic deformation and a full recovery of the foamlike layer of CCNTs under various impact velocities. 
\begin{figure}[ht]
\begin{center}
\includegraphics[width=7 cm]{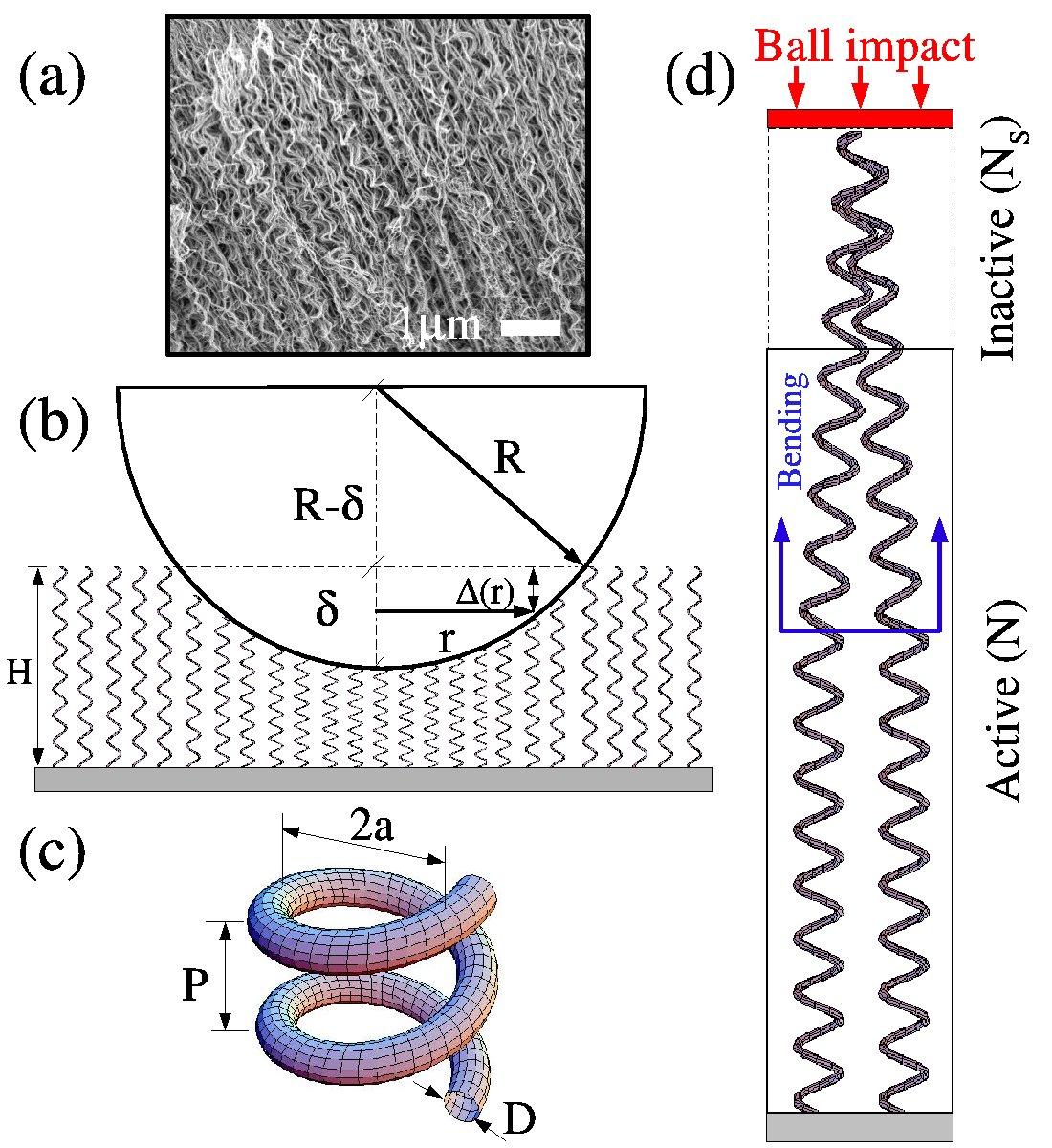}
\caption{(Color online) (a) Scanning electronic microscopy picture of a CCNT forest \cite{4}. (b) Schematic diagram of a CCNT forest deformed by a drop-ball. $H$ is the forest thickness, $R$ is the ball radius, and $\delta$ is the total depth displacement of the ball into the forest. $\Delta$ is the axial deformation of the spring at a distance $r$ from the center of the contact between the ball and the forest. (c) Parameters of a helical spring: $a$ is the coil radius, $P$ is the spring pitch and $D$ spring diameter. (d) Schematic representation of the entanglement between two adjacent coil bundles forming inactive, or ``solid'', and active turns. 
}
\label{fig1}
\end{center}
\end{figure}

The contact force exhibits a strongly nonlinear dependence on displacement and appears fundamentally different from the response of a forest of CNTs \cite{qi,mesarovic,1,2}. The obtained results in \cite{4} have been compared to the Hertz elastic model \cite{landau} of a solid sphere and a planar surface in contact where, if $F$ is the force of contact and $\delta$ is the displacement, $F\sim$ $\delta^{1.5}$ \cite{landau}. The results reported in Ref.~\cite{4} showed a nonlinear response of the CCNT forests having a force-displacement relation of $F\sim$ $\delta^{2.2}$, different than the Hertzian case. It has been hypothesized that sideways interactions of the compressed CCNTs can be associated with the strong nonlinear behavior in CCNT forests, but the physical mechanisms behind it are still unclear.

In this work we propose a model for the impact response of
a forest of CCNTs. The model takes into account: i) the individual elastic
contribution from each CCNT in contact with the drop-ball; ii) the
geometry of the surface of contact between the forest and the
drop-ball; and iii) sideways interactions of the compressed
CCNTs through an entanglement process. As discussed below we show that this model can describe the strongly nonlinear behavior observed in the recent experiments with CCNT forests~\cite{4}.

In the model the CCNT forest of thickness $H$ is considered as a set of identical springs that individually interact with the drop-ball and that may laterally interact with other springs. The drop-ball is considered to have a
rigid spherical surface of radius $R$. Figure \ref{fig1}(b) illustrates the penetration of the ball into the CCNT forest and the geometric features of the model. The total depth displacement of the ball into the CCNTs forest is
$\delta$. Each bundle of CCNTs is modeled as a helical spring of pitch $P$, and with a Hooke's constant $k=(GD^4)/(64a^3N)$~\cite{gjerde}, where $G$ is the shear modulus, $N$ is the number of active turns, $a$ is the coil radius, and $D$ is the bundle diameter (Fig. 1 (c)). Typical values from experiment are $H=$ 100 $\mu$m, $R=$ 1000 $\mu$m, $2a=$ 0.45 $\mu$m, $D=$ 0.1 $\mu$m, $\delta=$ 3 $\mu$m, and $P=$ 0.9 $\mu$m~\cite{4,bandaru}. We also considered the situation where the bundles of CCNTs are very close to each other. From Ref.~\cite{4} the total density of CCNTs in the forest is $\sim$100/$\mu$m$^2$. For a rough estimative, we obtained a separation between the centers of adjacent coils of $\sim$0.1 $\mu$m, assuming an ordered squared grid of CCNTs distributed on the surface with the value of the total density of CCNTs. This separation value is of the order of the estimated bundle diameter ($\sim$0.1 $\mu$m). Therefore, it is reasonable to assume the existence of a certain degree of entanglement between some parts of the CCNTs throughout the forest which would contribute to sideway interactions. Furthermore, it is probable that changes in the entanglement occur during the dynamic contact between the ball and the forest surface. This mechanism is illustrated in Fig. \ref{fig1}(d), where it is shown that the impact of the ball on the forest top surface would cause a bending of the tips of each CCNT, leading to contacts among the superior turns of the adjacent coils. The entanglement mechanism can be very complicated and a long-range process. In the present model, the effect of this mechanism in the forest response is translated into the reduction of the number of active turns of each CCNT, i.e.,
\begin{equation}
\label{NA}
N=N_T - N_S,
\end{equation}
where $N_T$ is the total number of turns of each CCNT
($N_T=H/P\cong110$), and $N_S$ is the number of turns that become
inactive due to the entanglement process (Fig. 1(d)). The inactive turns form a ``{\textit solid}'' phase that do not contribute to the elastic response of the CCNT. This approach has been used by Rodrigues {\textit et
al.}~\cite{6} to obtain a nonlinear relation between the force and
displacement of a conic spring. Based on the small displacement ($\sim$ 3 $\mu$m) of the top forest surface compared to the forest thickness ($\sim$ 100 $\mu$m), we also assume a short-range effect of the entanglement, i.e., only nearest neighboring coils interact. Initial entanglement prior to the impact can be incorporated in the initial number of active turns. The short-range interaction is included into the model by making the rate of increase of the number of inactive turns during the contact between the drop-ball and the forest surface proportional to the ball velocity:
\begin{equation}
\label{dNS}
\frac{dN_S}{dt}=\eta v ,
\end{equation}
where $\eta$ is a measure of how many turns become inactive per unit
length of the displacement. Despite the complexity of the interaction
that forms the entanglement, $\eta$ is considered here constant and
determined from experimental data \textit{a posteriori}.

Since $vdt=dz$, where $z$ is the direction perpendicular to the
forest surface (that is the direction of the movement of the
ball), Eq. (\ref{dNS}) can be easily integrated to give the following
expression for the number of active turns:
\begin{equation}
\label{NAfinal}
N(\Delta)=N_T-\eta \Delta,
\end{equation}
where $\Delta$ is the axial deformation of each CCNT. Figure
\ref{fig1}(b) shows how to determine the value of $\Delta$ as a
function of $\delta$ and the distance $r$ from the center of the
contact between the ball and the CCNTs forest (hereafter called
``center of contact'' for short). From geometrical considerations we
have
\begin{equation}
\label{deltaz}
\Delta(r)=\sqrt{R^2 - r^2}-(R-\delta).
\end{equation}

The total force $F$ consists of a summation of the forces of each
spring that interacts with the drop-ball:
\begin{equation}
\label{f1}
F=k\sum_{i=0}^{M}n_i\frac{\Delta(r_i)}
{1-(\eta/N_T)\Delta(r_i)}  \, ,
\end{equation}
where $i$ represents the set of $n_i$ springs that are at the same
distance $r_i$ from the center of contact. At the edge of the contact
area, $M$ is such that $\Delta(r_{M})=0$, i.e.,
\begin{equation}
\label{imax}
M=\frac{\sqrt{2R\delta-\delta^2}}{2a}.
\end{equation}
We assume a circular symmetry of the projection of the surface of
contact between the ball and the CCNTs forest, on the plane of the
forest. Therefore, $n_i$ is given by
\begin{equation}
\label{ni}
n_0=1 \,\,\,\, \mbox{and} \,\,\,\, n_i=\frac{2\pi r_i}{2r}=2\pi i \; ,
\,\, i=1,2,3,... M,
\end{equation}
where we used that $r_i=2ai$. The index $i=0$ is for the first coil hit by the ball. Equation (\ref{ni}) is, of course, an
approximation for the situation where the concentration of the CCNTs
are such that they are beside each other. 
Equation (\ref{ni}) will be a good estimative of the number of coils at
distance $r_i$ of the center of contact if the penetration of the ball
is enough to ensure that many coils are hit by the drop-ball. That
is the case considered in the experiment, where a total circular
contact area has a radius of $\sim$ 77 $\mu$m and the
CCNT bundle radius is $\sim$ 0.2 $\mu$m \cite{4}. For smaller depths
Eq. (\ref{ni}) should be corrected for a better result.

Since $\delta/R\ll1$ it is reasonable to use the following
approximations: $\Delta(r)\cong \delta-(2i^2r^2)/R$ and $M\cong
\sqrt{2R\delta}/(2a)$. Thus Eq. (\ref{f1}) can be rewritten as
\begin{equation}
\label{ff2}
F\cong \frac{k\delta}{1-(\eta/N_T)\delta}+2\pi
k\sum_{i=1}^{M}\frac{\displaystyle i (\delta R-2i^2a^2)}{\displaystyle
[1-(\eta /N_T)\delta]R+2\eta i^2a^2}.
\end{equation}
A simpler expression for $F$ cannot be derived from
Eq.~(\ref{ff2}). However, by neglecting the term $2 \eta \,i^2a^2$ from
the denominator, the following expression can be obtained
\begin{equation}
\label{Ffim}
\displaystyle F=k\frac{\pi R \delta^2}
{4 a^2 [1-(\eta /N_T)\delta]} \, ,
\end{equation}
where terms proportional to $\delta/R$ were also neglected.
%
\begin{figure}[ht]
\begin{center}
\includegraphics[width=7 cm]{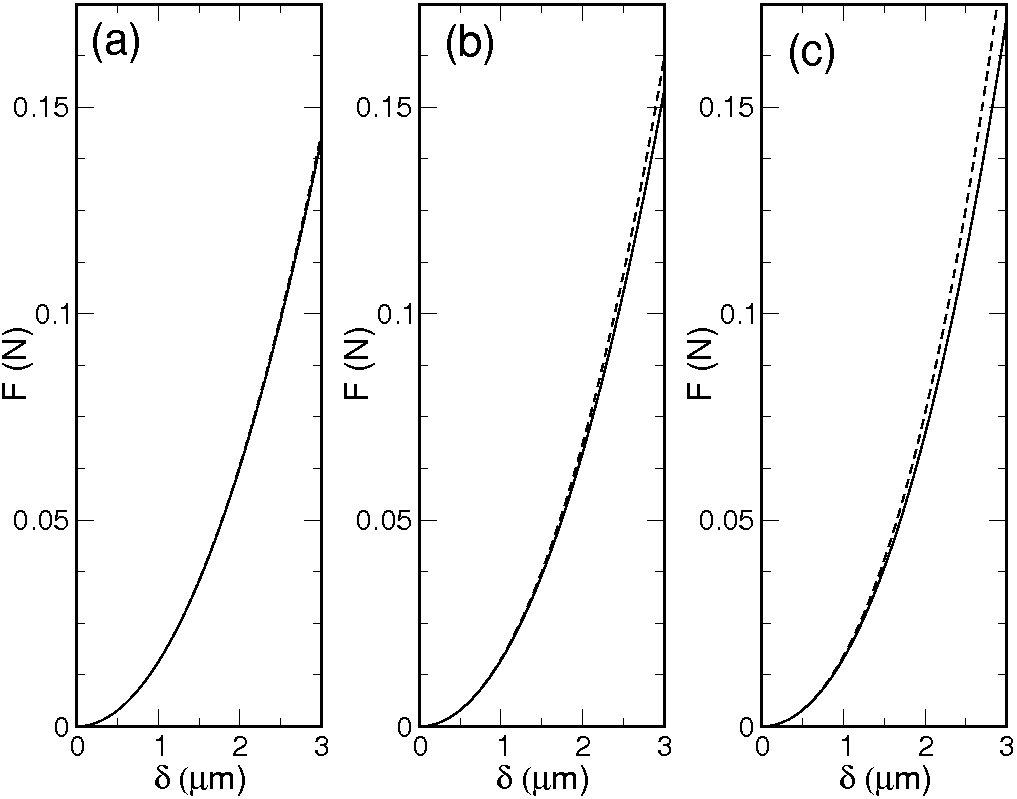}
\caption{Force-displacement curves obtained from Eq. (\ref{ff2}) when the 
term 2$\eta i^2a^2$ is present (solid lines, no approximation) or not
(dashed lines) for (a) $\eta =$1/$\mu$m, (b) $\eta =$5/$\mu$m, and (c)
$\eta =$10/$\mu$m. The following parameters where used: $a =$
0.225 $\mu$m, $R =$ 1000 $\mu$m, $N_T =$ 110, and $k =$1 N/m.}
\label{fig2}
\end{center}
\end{figure}

The result of these approximations to the final behavior of $F$ is shown in Fig. \ref{fig2} using typical experimental parameters. We can see that this approximation is physically sound for $\eta \lesssim$ 10/$\mu$m. We will see that this condition is satisfied in the experiments\cite{4}.

\begin{figure}[ht]
\begin{center}
\includegraphics[width=7 cm]{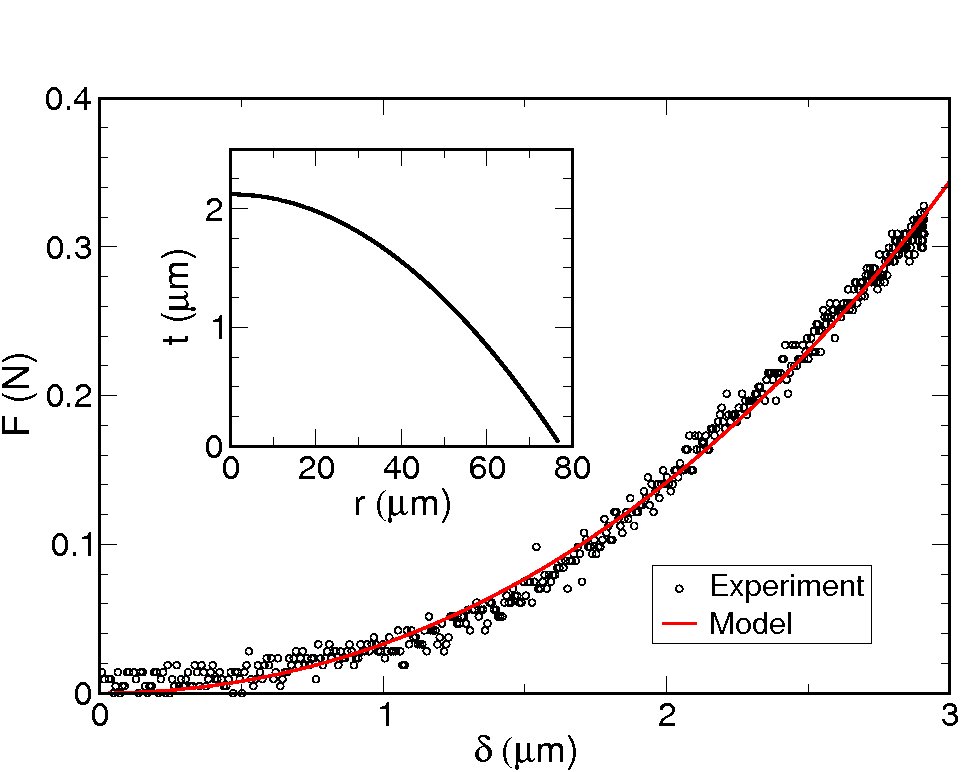}
\caption{(Color online) Behavior of the force as function of the displacemet 
of the drop-ball during contact with the CCNT forest. The model curve
was obtained using Eq. (\ref{Ffim}) with $k =$ 1.988 N/m and $\eta =$
7.066/$\mu$m. The inset graph shows the behavior of the thickness of
the entanglement as function of distance from the center of impact ($r=0$).}
\label{fig3}
\end{center}
\end{figure}

Using Eq. (\ref{Ffim}) to fit the experimental data~\cite{4} we obtain
$k=$ 1.988 N/m and $\eta =$ 7.066/$\mu$m. The comparison with
experiment is presented in Figure \ref{fig3}. 

The importance of the entanglement for the nonlinear behavior of the
forest of CCNTs can be tested by recalculating the total force with
$\eta =0$ in Eq. (\ref{NAfinal}), yielding to
\begin{equation}
\label{FN0}
F= k\, \frac{\displaystyle \pi R\delta^2}{\displaystyle
4a^2} \, .
\end{equation}
Eq. (\ref{FN0}) shows that the force $F$ is proportional to $\delta^2$. Even being close to the relation found in \cite{4}, this result still not capture the full nonlinear behavior of the impact response of a forest of
CCNTs. However, when the entanglement is turned on again, our model, with
the fitted parameters $k$ and $\eta$, recovers the previous
experimental fitting, $F=A\delta^{m}$, $A$ = 0.031 and $m$ = 2.2~\cite{4}, what shows that the entanglement formed by lateral deformations
of the CCNTs is necessary to explain the full strongly nonlinear behavior of the impact response of a drop-ball onto a forest of CCNTs.

In a loading experiment, Cheng \textit{et al.} have measured the spring constant value of a single amorphous carbon nanocoil in a low-strain regime (nanocoil elongation $\lesssim$3 $\mu$m) as being 0.12 N/m \cite{chen}. The number of turns of that nanocoil is about 10 which leads to $\sim$1.2 N/m for the value of the spring constant of a single turn. Using the derived $k$ value of the present model from experimental data ~\cite{4} we can estimate the spring constant of a single turn of an individual CCNT. According to Ref.~\cite{4} the bundle is formed by $\sim$25 nanocoils. Assuming that the bundle is formed by a parallel association of CCNTs, our estimative for the spring constant of a single CCNT is $k_s=k/25\cong0.08$ N/m. The spring constant of a single turn of the CCNT in a compression experiment is, then, $k_s\times N_T=8.7$ N/m, which is the same order of magnitude of the value for a single amorphous carbon nanocoil measured in a tension experiment. 

As previously mentioned, $\eta$ is the number of turns that become
inactive per unit length of displacement. Supposing that the
entanglement is formed by inactive turns that form a ``{\textit solid
}'' phase in the top of the CCNT forest, we can estimate the thickness,
$t$, of the entanglement at a given distance $r$ from the center of
impact as:
\begin{equation}
\label{thick}
t(r)=\eta D\Delta(r).
\end{equation}
where $\Delta(r)$ is given by Eq. (\ref{deltaz}). The inset graph of Fig. \ref{fig3} displays $t$ as a function of $r$. For the center of impact, the
thickness is 2.1 $\mu$m and the number of inactive turns of the coil
at the center of impact is $\sim$21, about 20\% of the total number of
turns of the total forest thickness. These results are compatible with what we expect for an entanglement formed by the superior parts of the CCNTs due to the impact of a drop-ball on the forest.

If, instead of a ball, we have an approximated perfect cube or
parallelepiped with a finite contact area $A$ with the forest, falling down on the forest of CCNTs, its impact response can be estimated using Eq.~(5). In this case, all nanocoils feel the same axial deformation $\delta$, and the
sum in Eq. (5) can easily performed to give $F \sim n_c \delta/[ 1 -
(\eta/N_T) \delta ]$, where $n_c$ is the number of nanocoils in contact
with the face of the cube or parallelepiped, and is simply given
by $n_c\sim A/(\pi a^2)$. The response force of the forest, then, will be
approximately linear with $\delta$, for small $\delta$ values, and start growing nonlinearly for larger $\delta$ ones, as consequence of the entanglement. 

It should be stressed the limit of validity of our model (represented by
Eq. (\ref{dNS})) takes into account the experimental condition \cite{4} of small forest deformations. One of the predictions of the model that can be experimentally tested is the thickness value of the entanglement of the top forest surface. We hope the present work will stimulate further experiments to test the validity of the present model. 

In conclusion, we have derived an analytical model for the nonlinear behavior of the impact response of a forest of CCNTs including geometrical and physical aspects during the forest compression. We showed that the nonlinear behavior is fully described when the entanglement of the coiled carbon nanotubes in the superior part of the forest surface is incorporated into the model. This entanglement among neighbors is due to the bending of the coil tips produced by the ball impact. Under the experimental conditions of small deformations \cite{4} the model predicts an entanglement thickness of $\sim$ 2 $\mu$m at the maximum forest compression. The model results point out to the importance of the coil entanglements for the elastic behavior of such systems. 
The present model is able to provide, by matching experimental values, estimates of the spring constant of a single CCNT and the level of entanglement between CCNTs. 
These aspects can play an essential role in the future design of micro-electro-mechanical systems devices, new shock protecting layers, and composites for microelectronic packaging and vibration mitigating materials where CCNT structural entanglement could be present.

We acknowledge the financial support from IMMP/MCT, IN/MCT, BNN/CNPq, THEO-NANO, and the Brazilian agencies FAPESP, CAPES and CNPq. AFF acknowledges CNPq scholarship.

\end{document}